\documentclass[showpacs,amsmath,amssymb,superscriptaddress]{revtex4}

\usepackage{graphicx}% Include figure files
\usepackage{dcolumn}% Align table columns on decimal point
\usepackage{bm}% bold math

\begin{document}

%\preprint{APS/123-QED}

\title{Phase diagram and critical behavior of the three-state majority-vote model}
\author{Diogo F. F. Melo}
\affiliation{Departamento de F\'{\i}sica,
Universidade Federal de Pernambuco, 50670-901, Recife-PE, Brazil}
\author{Luiz F. C. Pereira}
\affiliation{School of Physics, Trinity College Dublin, Dublin 2, Ireland}
\author{F. G. B. Moreira}
 \email{brady@df.ufpe.br}
\affiliation{Departamento de F\'{\i}sica,
Universidade Federal de Pernambuco, 50670-901, Recife-PE, Brazil}

\date{\today}% It is always \today, today,
             %  but any date may be explicitly specified

\begin{abstract}
The three-state majority-vote model with noise on Erd\"os-R\'enyi's random graphs has been studied.
Using Monte Carlo simulations we obtain the phase diagram, along with the critical exponents. Exact results for limiting cases are presented, and shown to be in agreement with numerical values. We find that the critical noise $q_{c}$ is an increasing function of the mean connectivity $z$ of the graph. The critical exponents $\beta / \bar{\nu}$,
$\gamma / \bar{\nu}$ and $1 / \bar{\nu}$ are calculated for several values of connectivity. We also study the globally connected network, which corresponds to the mean-field limit $z = N-1 \rightarrow \infty$.
Our numerical results indicate that the correlation length scales with the number of nodes in the graph, which is consistent with   an effective dimensionality equal to unity.
\end{abstract}

\pacs{64.60.Cn, 05.10.Ln, 64.60.Fr, 75.10.Hk}% PACS, the Physics and Astronomy
                             % Classification Scheme.
%\keywords{Suggested keywords}%Use showkeys class option if keyword
                              %display desired
%Valid PACS numbers may be entered using the \verb+\pacs{#1}+ command.
\maketitle

\section{\label{sec1}INTRODUCTION}

The majority-vote model with noise defined on a regular lattice is a system of spins where each one is allowed to be on two states only   \cite{marques,model91,model92,santos,mendes,kim,kwak}. In this two-state model, each spin assumes the state of the majority of its neighboring spins with  probability $(1-q)$ and the opposite state with probability $q$. The system presents an order-disorder phase transition as  the noise parameter $q$ reaches a critical value $q_c$.  Studies on the regular square lattice found $q_c = 0.075 \pm 0.005$ \cite{model92}, and critical exponents equal to those for the equilibrium Ising model in accordance with the conjecture by Grinstein {\it et al. } \cite{grinstein}.

The two-state majority-vote model (MV2) was also studied on a variety of complex networks \cite{paulo,felipe,welinton4,welinton2,welinton3,welinton1,holme}, such as undirected and directed random graphs \cite{erdos-renyi}, small-world networks \cite{ws}, and  Barab\'asi-Albert scale-free networks \cite{banet}. On undirected and directed random graphs, it was shown that $q_c$ is an increasing function of the connectivity of the graphs \cite{felipe,welinton1,welinton2}. On small-world networks, the critical noise is an increasing function of the rewiring probability \cite{paulo,welinton3}.
More general, these studies have shown that MV2 models defined on different complex networks belong to different universality classes and the calculated critical exponents depend on the topology of the complex network \cite{castellano,review,book}.
The generalization to a three-state majority vote model (MV3) on a regular square lattice was considered by \cite{tania2,tania1}, where the authors found $q_c = 0.117 \pm 0.001$. The resulting critical exponents for this non-equilibrium MV3 model are in agreement with the ones for the equilibrium three-state Potts model \cite{wu}, again supporting the conjecture of Ref. \cite{grinstein}.

In this paper we present an extensive study of the critical behavior of the three-state majority-vote model on Erd\"os-R\'enyi's  random graphs  \cite{erdos-renyi}.
Monte Carlo (MC) simulations and standard finite-size scaling techniques are used
to determine the critical noise parameter $q_{c}$, as well as the exponents
$\beta / \bar{\nu}$, $\gamma / \bar{\nu}$ and $1 / \bar{\nu}$ for several values of the mean
connectivity $z$ of the graph.  We also study the globally connected network case.
The phase diagram of the system is presented, and compared to our previously obtained diagram for the two-state model \cite{felipe}.
Exact results for quantities of interest are obtained for the limiting cases $q \rightarrow 0$, and $q \rightarrow 2/3$, which agree with the simulation results.

This work is organized in the following way: In section \ref{sec2} we describe
the non-equilibrium three-state majority-vote model with noise, and introduce
the relevant quantities used in our simulations. Sections \ref{sec3} and \ref{sec4} contain our results along with a discussion. Finally, in the last section we present our
conclusions.

\section{\label{sec2} Model and formalism}

The three-state majority-vote model with noise is defined by a set of
spin variables \{$\sigma_{i}$\}, where each spin is associated to one vertex of an Erd\"os-R\'enyi's random graph and can have the values $1, 2,$ and $3$. The connectivity of a vertex is defined as the total
number of bonds connected to it, that is $k_{i}=\sum_{j}c_{ij}$, where
$c_{ij}=1$ if there is a link between the sites $i$ and $j$ and $c_{ij}=0$
otherwise. A random graph is completely characterized by the mean number of
connections per site, {\it i.e.} the average connectivity $z$, and the total number of sites $N$.

The system evolves in time according to the following rules: For each spin we determine the state of the majority of its neighboring spins, that is, all the spins that are linked to it. With probability $(1-q)$  the new state of the spin agrees with the majority state of its neighbors and it disagrees with probability $q$, which is known as the noise parameter. In the case of a tie between the three possible states, each state is chosen with equal probability $1/3$.
In the case of a tie between two majority states, the spin assumes each one of these states with equal probability $(1-q)/2$, and the minority state with probability $q$. Finally, in the case of a single majority state, the two minority states occur with equal probability $q/2$, and the majority state with probability $(1-q)$.  It is clear that the rules just described present the $C_{3\nu}$ symmetry with respect to the simultaneous change of all states $\sigma$.

Let $k_i^{(\alpha)}$ be the number of neighbors of site $i$ in state $\alpha=1,2,3$, therefore $k_i^{(1)} + k_i^{(2)} + k_i^{(3)} = k_i$. According to the above rules we can write the following probabilities for a given spin to assume the state $1$:
\begin{equation}\label{rules}
\begin{array}{l}
P(1 | k_i^{(1)} = k_i^{(2)} = k_i^{(3)} )  =  1/3 \\
P(1 | k_i^{(1)} = k_i^{(2)} > k_i^{(3)} )  =  (1-q)/2 \\
P(1 | k_i^{(1)} < k_i^{(2)} = k_i^{(3)} )  =  q \\
P(1 | k_i^{(1)} > k_i^{(2)}, k_i^{(3)} )  =  1-q \\
P(1 | k_i^{(1)}, k_i^{(2)} < k_i^{(3)} )  =  q/2.\\
\end{array}
\end{equation}

The probabilities for the other two states are obtained by the symmetry operations of the $C_{3\nu}$ group.
For example, let us consider a neighborhood corresponding to the fourth and fifth rules, where we have a single majority state. In this case, the two minority states (say, states 2 and 3) occur with equal probability $q/2$, and the majority state with probability $(1 - q)$. We can write
$P(1| k(1) > k(2) , k(3))  = 1-q$ from the fourth rule, and
$P(2| k(2), k(3) < k(1))  = P(3| k(3), k(2) < k(1))  =  q/2$, from the fifth rule.
It is worth mentioned that the condition $1 - q$ (= Probability of choosing the majority state 1) $> q/2$ (= Probability of choosing a minority state, either 2 or 3) is valid for $q < 2/3$, and we conclude that $q=2/3$ is the limit value for the noise parameter in the present three state MV model. Moreover, the probabilities defined by Eq.(\ref{rules})  satisfy $P(1|\dots) + P(2|\dots) + P(3|\dots) =1$.

To study the critical behavior of the model we consider the magnetization
$M_{N}$, the susceptibility $\chi_{N}$, and the Binder's fourth-order
cumulant $U_{N}$. These quantities are defined by
\begin{equation} \label{eq:mag}
M_{N}(q) = \left\langle \left\langle m \right\rangle _{t} \right\rangle _{c}
\end{equation}
\begin{equation} \label{eq:sus}
\chi_{N}(q) = N\left[ \langle ~ \langle m^{2}\rangle _{t} \rangle _{c} - \langle
\langle m \rangle _{t} \rangle _{c}^{2}\right]
\end{equation}
\begin{equation} \label{eq:cumu}
U_{N}(q) = 1 - \frac{\langle ~ \langle m^4 \rangle _{t} \rangle _{c} }{3\langle ~
\langle m^2 \rangle _{t} \rangle _{c}^2},
\end{equation}
where N is the number of vertices of the random graph with fixed $z$,  $\langle ... \rangle_{t}$ denotes
time averages taken in the stationary regime, and $\langle ... \rangle_{c}$
stands for configurational averages. In Eqs. (\ref{eq:mag})-(\ref{eq:cumu}) $m$ is defined in analogy to the magnetization in the three-state Potts model as the modulus of the magnetization vector, that is $m=(m_1^2 + m_2^2 +m_3^2)^{1/2}$, whose components are given by
\begin{equation} \label{eq:magcomp}
m_{\alpha} = \sqrt{\frac{3}{2}} \left[ \frac{1}{N} \sum_i^{} \delta( \alpha, \sigma_i) - \frac{1}{3} \right],
\end{equation}
where the sum is over all sites in the graph, $\delta(\alpha, \sigma_i)$ is the Kronecker delta function, and we introduce the factor $\sqrt{3/2}$ in order to normalize the magnetization vector.

In the critical region we assume the following finite-size scaling (FSS) relations \cite{park}
\begin{equation} \label{mFSS}
M_{N}(q) = N^{-\beta / \bar{\nu}}\tilde{M}(\varepsilon N^{1 / \bar{\nu}})
\end{equation}
\begin{equation} \label{xFSS}
\chi_{N}(q) = N^{\gamma / \bar{\nu}}\tilde{\chi}(\varepsilon N^{1 / \bar{\nu}})
\end{equation}
\begin{equation} \label{uFSS}
U_{N}(q) = \tilde{U}(\varepsilon N^{1 / \bar{\nu}})
\end{equation}
where $\varepsilon = q - q_{c}$, and the universal scaling functions $\widetilde{M}$, $\widetilde{\chi}$ and $\widetilde{U}$ only depend on the scaled variable $ x=\varepsilon N^{1/\bar{\nu}}$.
The above FSS relations follow from the {\it ansatz} that the correlation length scales with the number of nodes in the graph, that is $\xi \sim N$, which is consistent with an effective dimensionality equal to unity.

From the size dependence of $M_{N}$ and $\chi_{N}$ we can obtain the exponents $\beta / \bar{\nu}$ and $\gamma / \bar{\nu}$, respectively.
The correlation length exponent $\bar{\nu}$ is calculated from the size dependence of the derivative of Binder's fourth-order cumulant with respect to the noise parameter, $U_{N}^{\prime}(q=q_c)$.
Furthermore, we use the hyperscaling relation
\begin{equation} \label{hsc}
2\beta / \bar{\nu} + \gamma / \bar{\nu} = D_{eff},
\end{equation}
to estimate the effective dimensionality of the system $D_{eff}$ in order to check the FSS prediction for an effective dimensionality equal to unity.

\section{\label{sec3} Results}

\subsection{Exact Results}

First we notice that there are only two independent components of the magnetization vector $m_{\alpha}$, since they obey the relation
\begin{equation}
m_1 + m_2 + m_3 = 0.
\end{equation}
Moreover, the norm of the magnetization is invariant with respect to any $C_{3 \nu}$ group symmetry operation.

Let us consider the limit $q \rightarrow 0$, where the probability of agreeing with the majority state equals unity. In this situation, after a transient period the system reaches the ordered steady state. Without loss of generality we can assume $\sigma_i=1$ for all sites. In this case $m_1 = \sqrt{2/3}$ and $m_2 = m_3 = -1/\sqrt{6}$, thus $M_N(0) = 1$. It is also straightforward to check that $\chi_N(0) = 0$, and $U_N(0) = 2/3$.

In the opposite limit ($q \rightarrow 2/3$) the system reaches a disordered steady state, where the average of each component of the magnetization vanishes. In fact, in this limit the probability of a given spin agreeing with the majority of its neighbors equals the probability of it agreeing with any of the other two minority states, {\it i.e.} $(1-q)=q/2$.
It is possible to write the probability distribution for the order parameter as a Gaussian distribution in the form \cite{tania2}
\begin{equation}
P(m) = \frac{a}{\pi} e^{-a m^2},
\end{equation}
where $a=1/ \langle m^2 \rangle$.
From this distribution we obtain that $M_N(q \rightarrow 2/3) = \sqrt{\pi}/2 \langle m^2 \rangle^{1/2}$, $\chi_N(q \rightarrow 2/3) = N(4/ \pi -1) M_N^2$, and $U_N(q \rightarrow 2/3) = 1/3$. Moreover, it follows that $\langle m^2 \rangle \sim N^{-1}$,
$M_N(q \rightarrow 2/3) \sim N^{-1/2}$, and $\chi_N(q \rightarrow 2/3) \sim N^{0}$.
These exact results as well as the predict size $N$-dependence of the relevant quantities are in agreement with numerical results from simulations for all networks considered.

\subsection{Simulation}
We begin our simulations generating a random graph of size $N$ and mean connectivity $z$ in a disordered configuration where the state of each spin is $1, 2$ or $3$ with the same probability. We used systems of size $N=1000, 2000,4000, 6000, 10000, 50000$, and varied $z$ from $1$ to $50$. To perform the dynamics we choose a site at random and, for a given fixed value of the noise parameter, we update its state in accordance with the dynamics rules given by Eq. (\ref{rules}). A Monte Carlo step (MCS) is defined as $N$ updates. We waited $N_{r}$ MCS needed for the system to reach the steady state, and the time averages, $\langle ... \rangle_{t}$, were estimated from the next $N_{s}$ MCS. The values of MCS used vary with $N$, $z$ and $q$, typically we used $N_{r} > N_{s} > 5000$ MCS. For all sets of parameters, we have generated at least $100$ distinct random networks in order to calculate the configurational averages $\langle ... \rangle_{C}$.

\begin{figure}
\includegraphics[width=0.35\textwidth,angle=-90]{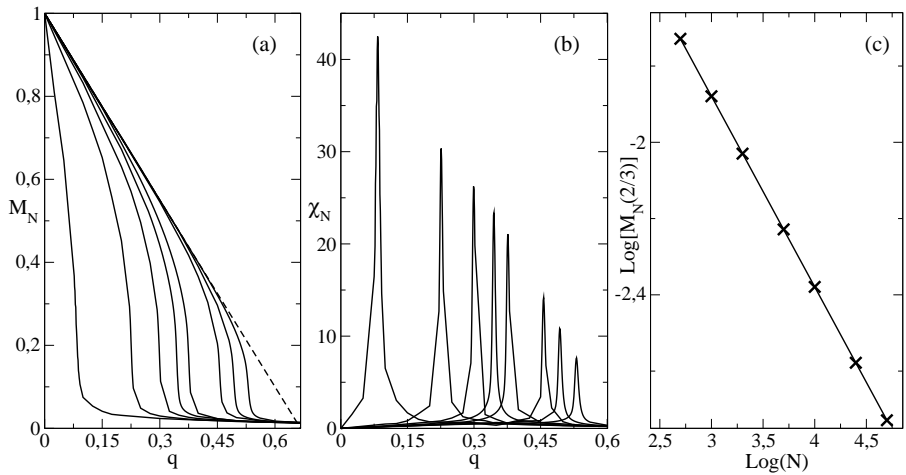}

\caption{\label{mag-susc} Dependence of the magnetization (a) and the susceptibility (b) on the
 noise parameter $q$, for $N=4000$ nodes. From left to right we have
$z=2$, $4$, $6$, $8$, $10$, $20$, $30$, and $50$. In (a) the dashed line corresponds to the mean-field result $m=-\frac{3}{2}(q - \frac{2}{3})$, in the thermodynamical limit $z = N -1 \rightarrow \infty$. In part (c) we plot the size dependence of the magnetization at $q_c=2/3$ for the case of $z=N-1$, the globally connected network.}
\end{figure}

Fig. \ref{mag-susc} shows the magnetization $M_{N}$ and the susceptibility $\chi_{N}$ as functions of the noise parameter. The data were obtained from simulations on random graphs with $N=4000$ sites and several values of the average connectivity $z$.
In part (a) each curve for $M_{N}$, for a given value of $z$, clearly indicates
that there exists a phase transition from an ordered state to a disordered
state where the magnetization vanishes.
We also notice that the transition occurs at a value of the critical
noise parameter $q_{c}$, which is an increasing function of the mean connectivity $z$ of the random graph.
In part (b) we show the corresponding behavior of the susceptibility $\chi_{N}$. The value of $q$ where $\chi_{N}$
has a maximum is here identified as $q_{c}(N)$.

We also perform simulations for globally connected networks, that is, for random graphs with $N$ nodes and connectivity $z=N-1$. In the thermodynamical limit $z = N -1 \rightarrow \infty$, the magnetization is given by
$m=-\frac{3}{2}(q - \frac{2}{3})$ (the dashed line in Fig. 1(a)), from which we obtain the mean-field values $q_{c}^{(MF)}=\frac{2}{3}$ and $\beta=1$.
Fig. 1(c)  shows the dependence of the mean-field magnetization at $q_{c}^{(MF)}$ on the system size. The straight line confirms the scaling relation given by  $M_N(q = 2/3) \sim N^{-1/2}$. The slope of the resulting straight line yields the exponent $\beta / \bar{\nu}=1/2$. A similar analysis for the  susceptibility at $q_{c}^{(MF)}$ yields $\chi_N(q = 2/3) \sim N^{0}$, that is, $\gamma / \bar{\nu}=0$. Note that the critical behavior for the globally connected network is in agreement with the exact results discussed above in the limit case of $q = 2/3$. The  mean-field results for the critical noise parameter and critical exponents are given in Table \ref{tab1}.

\begin{figure}
\includegraphics[width=0.4\textwidth,angle=-90]{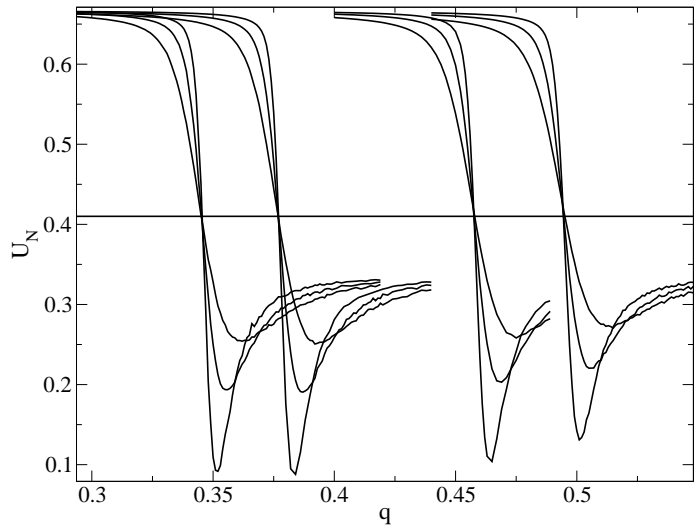}
\caption{\label{cumul} Binder's fourth-order cumulant as a function of $q$, for system sizes $N=1000,2000,4000$.
From left to right we have $z=8$, $10$, $20$, and $30$. The horizontal line indicates the critical value $U^{\star}=0.42$. The mean-field value is $U^{\star}=1/3$.}
\end{figure}

\begin{figure}
\includegraphics[width=0.4\textwidth,angle=-90]{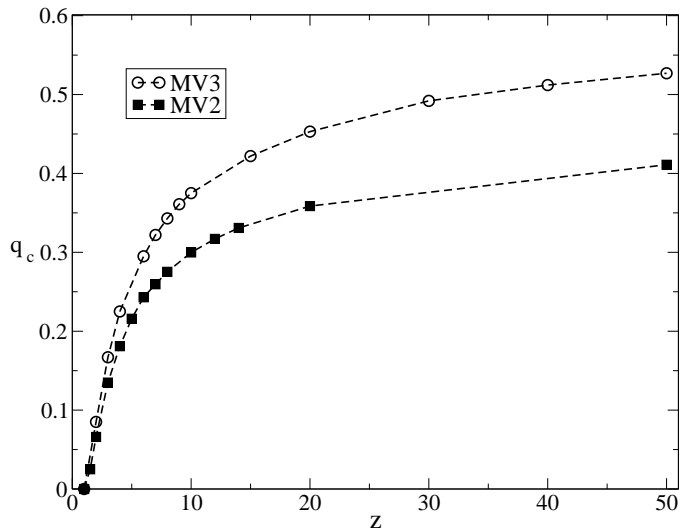}
\caption{\label{phased} The phase diagram for the three-state majority-vote model (this work),
showing the dependence of the critical noise parameter $q_{c}$ on the average connectivity $z$ of the random graph.
The phase diagram for the two-state model (Ref. \cite{felipe}) is also included for comparison.}
\end{figure}

In Fig. \ref{cumul} we plot Binder's fourth-order cumulant $U_{N}$
for different system sizes $N$ and four distinct values of $z$.
The critical noise parameter $q_{c}$, for a given value of $z$,
is estimated as the point where the curves for different values of $N$
intercept each other. We also obtained $U^\star=0.42(2)$ for the critical value of the cumulant at $q_{c}$,
which is independent of the connectivity $z$ of the graph. The dependence of $q_{c}$ on $z$ yields
the phase diagram for the MV3 model shown in Fig. \ref{phased}.

The phase diagram of the MV3 model on random graphs shows that for a given graph (fixed $z$) the system becomes ordered for $q<q_c$, whereas it has zero magnetization for $q \ge q_c$.
We notice that the increase of $q_{c}$ is more pronounced for small values of $z$. The error bars in $q_{c}$ (see Table \ref{tab1}) are much smaller than the symbols.
In the figure, it is also included the corresponding phase diagram for the MV2 model obtained from Monte Carlo
simulation in our previous work \cite{felipe}. For both models the system exhibits an ordered state
for all values of the mean connectivity greater than one. This is in agreement with the limiting value of $z=1$
for the existence of a percolating cluster and, therefore, the onset of long-range order in the system.
However, when $z \rightarrow \infty$ we obtain the upper limits $q_c=0.5$ and $q_c=2/3$,  for the MV2 and MV3 models respectively.

\begin{figure}
\includegraphics[width=0.4\textwidth,angle=-90]{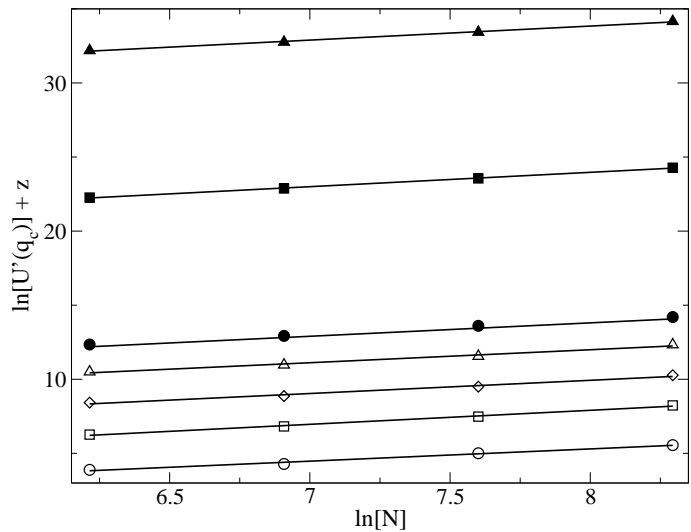}
\caption{\label{nu} (a) Plot of $\ln (U^{\prime}(q_{c})) + z$ vs $\ln N$. The exponent $1 / \bar{\nu}$, for a given $z$,
corresponds to the slope of the straight line obtained from a linear fit to the data.
From bottom to top, $z=2, 4, 6, 8, 10, 20, 30$.}
\end{figure}

Fig. \ref{nu} shows the dependence of the derivative of the Binder's fourth-order cumulant at $q=q_c$ on system size. For clarity we have added in each curve the respective value of $z$. The straight lines, obtained from simulations with different values of the mean connectivity $z$, confirm the scaling relation given by Eq. (\ref{uFSS}). For fixed $z$, the slope of the resulting straight line equals the exponent $1 / \bar{\nu}$. The results displayed in Table \ref{tab1} indicate a weak dependence of the correlation length exponent with $z$.

In order to obtain independent estimations for the critical noise parameter $q_{c}(z)$, as well as the exponent ratios $\beta / \bar{\nu}$ and $\gamma / \bar{\nu}$, for different values of the mean connectivity $z$, we consider the functions $\Phi$ and $\Psi$ defined as
\begin{equation}\label{phi}
\Phi_{N_1,N_2} = -b^{-1} \ln \frac{ M_{N_2}}{M_{N_1}}
\end{equation}
\begin{equation}\label{psi}
\Psi_{N_1,N_2} = b^{-1} \ln \frac{\chi_{N_2}}{\chi_{N_1}}
\end{equation}
where $b=\ln(N_2 / N_1)$. The above functions relate the magnetizations and susceptibilities calculated with two different system sizes, $N_1$ and $N_2$.  In fact, substituting the finite-size relation  (\ref{mFSS}) into Eq. (\ref{phi}),  we obtain
\begin{equation}
\Phi_{N_1,N_2} = \beta/\bar{\nu}  -  b^{-1} \ln \frac{ \tilde{M}(\varepsilon N_2^{1 / \bar{\nu}}) }{ \tilde{M}(\varepsilon N_1^{1 / \bar{\nu}})}.
\end{equation}
At the critical value $q_{c}(z)$, the last term vanishes and we obtain $\Phi(q_c) = \beta / \bar{\nu}$. Analogously, substituting the finite-size relation (\ref{xFSS}) into Eq. (\ref{psi}), we have $\Psi(q_c) = \gamma / \bar{\nu}$.

\begin{figure}
\includegraphics[width=0.5\textwidth,angle=-90]{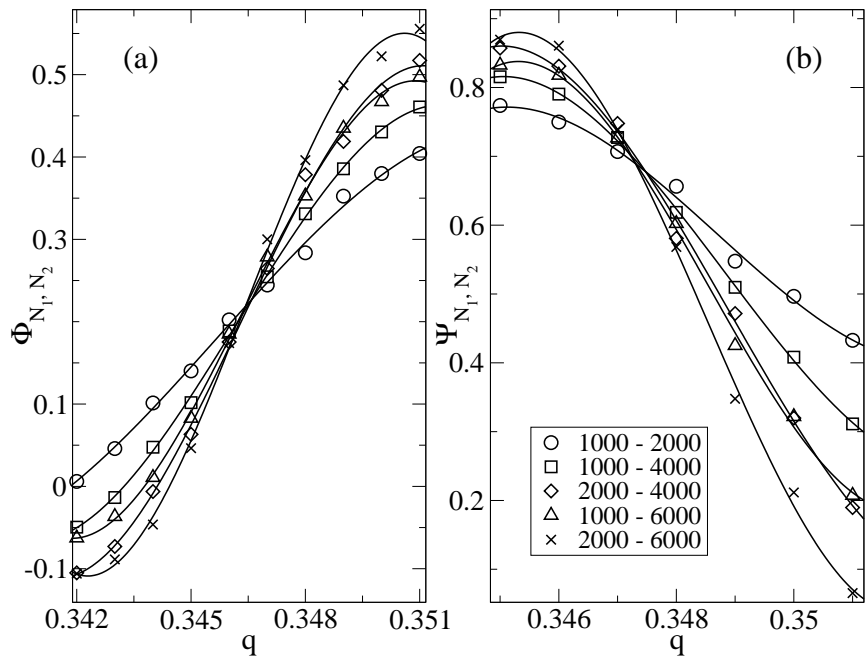}
\caption{\label{phipsi} The functions $\Phi$ and $\Psi$ for the case of mean connectivity $z=8$,
and five different pairs of $N_1$ and $N_2$. The intersection points give independent estimations
for $\beta / \bar{\nu}$, $\gamma / \bar{\nu}$, and $q_c$. The curves are cubic polynomial fitting to the data.}
\end{figure}

In Fig. \ref{phipsi} we show the dependence with noise of the functions $\Phi$ and $\Psi$, for $z=8$ and several values of $N_1$ and $N_2$. From the intersection points in Fig. \ref{phipsi}(a) and Fig. \ref{phipsi}(b) we obtain independent estimations for the critical noise parameter $q_{c}(z)$.
For all networks considered in the simulations, we have obtained a quite satisfactory agreement between the two values of $q_{c}(z)$ determined in this way and the corresponding ones that follow from the analysis of Binder's cumulant (Fig. \ref{cumul}).
Moreover, we used the relations $\Phi(q_c) = \beta / \bar{\nu}$ and $\Psi(q_c) = \gamma / \bar{\nu}$ to calculate the exponent ratios for different values of the mean connectivity $z$.
Table \ref{tab1} shows the results for different values of the mean connectivity $z$. 
We call the readers attention to the difference between the calculated values of the exponents in the case of $z=2$ and the corresponding results with $z>2$. This might be an indication that, for $z=2$, we need to take into account logarithmic corrections to the finite-size scaling relations \cite{caracciolo}.

\begin{figure}
\includegraphics[width=0.45\textwidth,angle=-90]{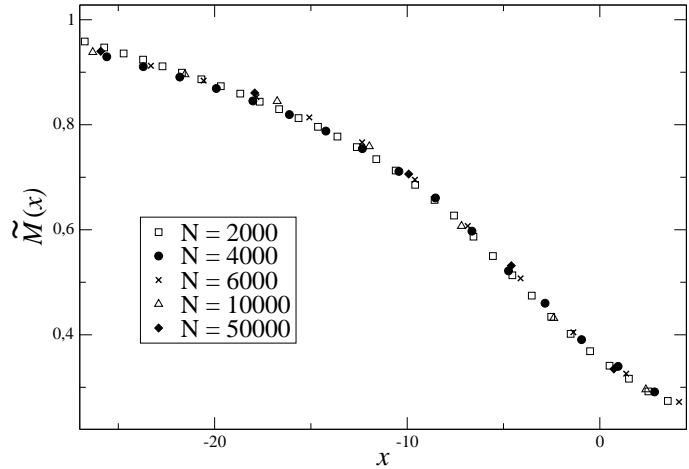}
\caption{\label{Mdatacol} The universal scaling function $\widetilde{M}$ versus the scaled variable $x=\varepsilon N^{1/\bar{\nu}}$.
Data-collapsing for five different values of $N$, with $z=8$.}
\end{figure}

Fig.~\ref{Mdatacol} shows the data-collapse plot for
$\tilde{M}(x)=M_{N}(q) N^{\beta / \bar{\nu}}$, which is a universal function of
the combined variable $x=N^{1 / \bar{\nu}} (q-q_c)$. We have also obtained quite good
data-collapse for $\tilde{\chi}(x)=\chi_{N}(q) N^{-\gamma / \bar{\nu}}$.
The collapsing of curves for five different system sizes corroborates the quoted values for $q_c$, $\beta / \bar{\nu}$, $\gamma / \bar{\nu}$ and $1 / \bar{\nu}$.

\begin{figure}
\includegraphics[width=0.45\textwidth,angle=-90]{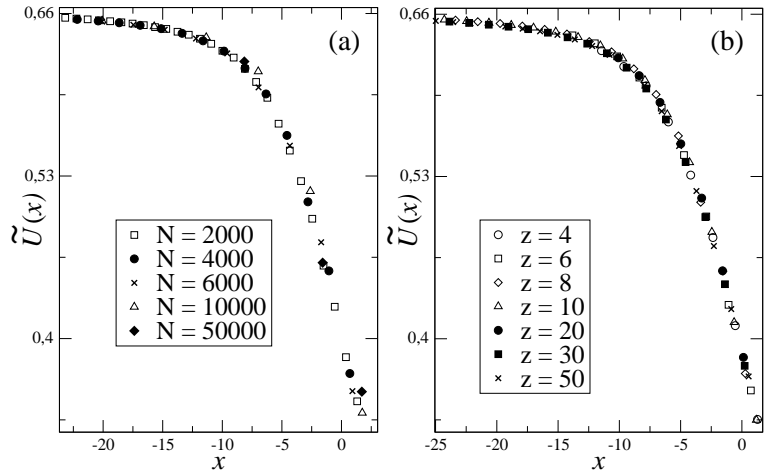}
\caption{\label{Udatacol}
Data-collapsing of the Binder's cumulant. In (a) we show the collapse for five different values of $N$, with $z=8$ fixed.
In (b) the collapse is obtained for seven different values of $z$, with $N=4000$ fixed.}
\end{figure}

On Fig. \ref{Udatacol} we present two different ways to obtain the data-collapsing for the universal scaling function $\widetilde{U}(x)$. Part (a) shows the standard data-collapse that follows from simulations with different values of system size $N$, for the case of mean connectivity $z=8$ fixed. In part (b) we have fixed $N=4000$ and used the data from simulations for varying connectivity. It is worth mentioning that in the last case the collapse for different values of $z$ was obtained by using the values for the critical parameter $q_c$ and the exponent $1/\bar{\nu}$ for the corresponding value of $z$ (see Table \ref{tab1}).
In general, the universal scaling functions only depend on the scaled variable $ x=\varepsilon N^{1/\bar{\nu}}$ for a given system. However, the two data-collapsing in Fig. \ref{Udatacol} indicate that $\widetilde{U}$ does not depend on the specific value of $z$, contrary to what is observed for $\widetilde{M}$ and $\widetilde{\chi}$.

\begin{table}[h]
\caption{\label{tab1} The critical noise $q_c$, the critical exponents, and the effective dimensionality $D_{eff}$, for the MV3 model on random networks with mean connectivity $z$. The mean-field (MF) results are shown in the last line.}
\begin{ruledtabular}
\begin{tabular}{rlllll}
z & $q_{c}$ & $\beta / \bar{\nu}$ & $\gamma / \bar{\nu}$ & $1 / \bar{\nu}$ & $D_{eff}$ \\
\hline
2 & 0.084(5) & 0.15(3) & 0.75(2)& 0.81(6) & 1.05(8) \\
4 & 0.228(1) & 0.20(1) & 0.65(1)& 0.94(5) & 1.05(3) \\
6 & 0.3015(5) & 0.198(5) & 0.66(5)& 0.89(7) & 1.06(6) \\
8 & 0.3458(2) & 0.2105(5) & 0.62(5) & 0.895(5) & 1.04(9) \\
10 & 0.3785(2) & 0.22(2) & 0.68(2) & 0.90(2) & 1.12(6) \\
20 & 0.4586(5) & 0.205(5) & 0.67(2) & 0.95(1) & 1.07(3) \\
30 & 0.4957(1) & 0.22(1) & 0.66(2) & 0.95(2) & 1.1(3) \\
50 & 0.533(1) & 0.22(2) & 0.64(7) & 0.92(2) & 1.08(6) \\
MF & 2/3      & 1/2     & 0       &  1/2    & 1        \\
\end{tabular}
\end{ruledtabular}
\end{table}

\section{\label{sec4} Discussion}

Table \ref{tab1} summarizes the values (along with errors) of the critical noise parameter
$q_{c}$, the critical exponents $\beta / \bar{\nu}$, $\gamma / \bar{\nu}$ and $1 / \bar{\nu}$, and the effective dimensionality of the system.
For all $z$ considered, including the mean-field limit $z = N-1 \rightarrow \infty$, the value $D_{eff} \simeq 1$ follows from the hyperscaling relation (Eq. (\ref{hsc})), in agreement with the scaling {\it ansatz} for the magnetization and susceptibility Eqs. (\ref{mFSS}, \ref{xFSS}). In fact, since our original work \cite{felipe} several authors have studied different spin models on varied complex networks, always finding an effective dimensionality equal to one \cite{welinton1,welinton2,welinton3,welinton4,holme}. Even though this result seemed surprising at first, it is a direct consequence of the scaling for the correlation length $\xi \sim N^{1/D_{eff}}$, with $D_{eff}=1$.

There are no previous works studying the three-state majority-vote model on Erd\"os-R\'enyi's graphs, to allow a direct
comparison of the present results. Yet, for completeness, it would be of interest to mention earlier simulations of the
majority-vote model on other kinds of networks. The only works on the MV3 model to this date, considered the model on a regular square lattice \cite{tania1,tania2}. They find the universal value of Binder's fourth-order cumulant to be $U^{\star} \simeq 0.61$, and the exponents $\beta/\bar{\nu}=0.134(5)$ and $\gamma/\bar{\nu}=1.74(2)$, all of which are in agreement with the results for the equilibrium three-state Potts model \cite{wu}. The present simulations of the MV3 model on globally connected networks yielded $U^{\star}=1/3$, and the following mean-field exponents: $\beta/\bar{\nu}=1/2$, $\gamma/\bar{\nu}=0$, $\beta=1$, $\bar{\nu}=2$, and $\gamma=0$.
From our simulation results we can conclude that the MV3 model defined on a regular square lattice, on Erd\"os-R\'enyi's random graphs, and on the  corresponding globally connected network (the mean-field limit) belong to different universality classes. 

Comparing the current results with the ones previously obtained for the two-state model on random graphs \cite{felipe}, we first notice that the ordered region ($q < q_c$) in the phase diagram of Fig. \ref{phased} is larger for MV3 than for MV2. This is expected since now we might obtain three possible majority states. It should also be clear from the calculated exponents that these models do not belong to the same universality class. In particular, $U^{\star} \simeq 0.30$ for MV2, and $U^{\star} \simeq 0.42$ for MV3.

\section{Conclusion}

We have obtained the phase diagram and critical exponents of the three-state majority-vote model
with noise on random graphs. The second-order phase transition which occurs in the model with mean connectivity $z>1$ has exponents that show a slight variation along the critical line. Nevertheless, our Monte Carlo simulations provide an effective dimensionality $D_{eff}$ equal to one for all values of $z$. This result, which is in agreement with several previous studies on spin models defined on complex networks, was shown to be a consequence of the {\it ansatz} that the correlation length scales with the number of nodes. Future work on the two- and three-state Potts model on random graphs would be of interest in order to provide a direct comparison with our results in light of the conjecture by Grinstein {\it et al.}, which states that reversible and irreversible models with same symmetry belong to the same universality class.

\begin{acknowledgments}
D.F.F. Melo is supported by CNPq. L.F.C. Pereira is supported by Science Foundation Ireland. We acknowledge partial
support from CNPq, FINEP and FACEPE.
\end{acknowledgments}

%\newpage %Just because of unusual number of tables stacked at end
%\bibliography{apssamp}% Produces the bibliography via BibTeX.

\end{document}